\documentclass[twocolumn,aps,showpacs,prb,tightenlines,amsmath,amssymb]{revtex4}
\usepackage{bm}
\usepackage{graphicx}              
\usepackage{amssymb}               
\usepackage{amsmath}                     
\usepackage{colordvi}
\usepackage{calrsfs} \newcommand{\bgreek}[1]{\mbox{\boldmath$#1$\unboldmath}}
\begin{document}   

\title{Hot-electron effect in spin relaxation of electrically injected electrons
  in intrinsic Germanium}
\author{T. Yu}
\author{M. W. Wu}
\thanks{Author to whom correspondence should be addressed}
\email{mwwu@ustc.edu.cn.}
\affiliation{Hefei National Laboratory for Physical Sciences at
  Microscale,  Key Laboratory of Strongly-Coupled Quantum Matter Physics
 and Department of Physics, 
University of Science and Technology of China, Hefei,
  Anhui, 230026, China} 
\date{\today}

\begin{abstract}
The hot-electron effect in the spin relaxation of electrically injected
electrons in intrinsic Germanium is investigated
 by the kinetic
  spin Bloch equations both analytically and
 numerically.  It is shown that in the weak-electric-field regime with $E\lesssim 0.5$~kV/cm, our
calculations has reasonable agreement with the recent
transport experiment in the hot-electron spin-injection configuration
 [Phys. Rev. Lett. {\bf 111}, 257204 (2013)]. We reveal that the spin
  relaxation is significantly enhanced at low temperature in the presence of
  weak electric field $E\lesssim 50$~V/cm, which originates from
  the obvious center-of-mass drift effect due to the weak electron-phonon
  interaction, whereas the hot-electron effect is demonstrated to be
  less important. This can explain the 
discrepancy between the experimental observation and the previous 
theoretical calculation [Phys.
Rev. B {\bf 86}, 085202 (2012)], which deviates from the experimental
results by about two orders of magnitude at low temperature.
 It is further shown that in the strong-electric-field regime
  with $0.5\lesssim E \lesssim 2$~kV/cm, the spin relaxation is enhanced due to the
  hot-electron effect, whereas the drift
effect is demonstrated to be marginal. Finally, we find that when $1.4 \lesssim E\lesssim 2$~kV/cm
which lies in the strong-electric-field regime,
 a small fraction of electrons ($\lesssim 5\%$) can be driven
  from the L to $\Gamma$ valley, and the spin relaxation rates are
  the same for the $\Gamma$ and L valleys in the intrinsic sample without impurity.  
 With the negligible influence of the spin dynamics in
  the $\Gamma$ valley to the whole system, the spin dynamics in
  the L valley can be measured from the $\Gamma$ valley by the 
standard direct optical transition method.

\end{abstract}


\maketitle 

\section{Introduction} 
Recently, growing attention has been paid to the electron spin dynamics in
Germanium (Ge)
 due to the series of experiments including both the
 electrical\cite{Loren6,Loren7,Loren8,Loren9,Loren10,Loren11,three-four,Li-Ge4} and optical
 measurements.\cite{optical1,optical2,Pezzoli-Ge1,Pezzoli-energy,Guite,optical3}
These experimental progresses pave the way to
 utilize the excellent property of Ge in the spintronic
 application.\cite{Loren-Ge1,Loren-Ge2,Pezzoli-Ge1,
Rioux-Ge1,Li-Si1,Tang-Ge1,Li-Ge2,Loren11,Jain-Ge1,Li-Ge3,Loren8}
 In Ge, due to the lack of the
inversion asymmetry, the D'yakonov-Perel' (DP) mechanism\cite{DP} is
absent. Moreover, the 
hyperfine interaction with the nuclei can be suppressed by the isotopic
purification.\cite{purification1,purification2} 
  Therefore, its spin relaxation time (SRT) is expected to be very 
long.\cite{Loren-Ge1,Loren-Ge2,Pezzoli-Ge1,
Rioux-Ge1,Li-Si1,Tang-Ge1,Li-Ge2,Loren11,Jain-Ge1,Li-Ge3,Loren8}
Furthermore, in
intrinsic Ge, the spin relaxation of electrically/optically injected electrons
 due to the electron-impurity scattering can be
further suppressed.\cite{Song_impurity}
 Therefore, with the mature nanoelectronic fabrication
 technology of the Group IV semiconductors, intrinsic Ge is promising for the design and
 development of spintronic
 devices.

The electrical methods with the application of electric
 and/or magnetic
 fields include the three- or four-terminal Hanle 
 configuration\cite{Loren6,Loren7,Loren8,Loren9,Loren10,Loren11,three-four} and
 the ballistic 
 spin-injection configuration.\cite{Li-Ge4}
 In the three- or four-terminal Hanle configuration, the
 typical SRTs measured are in the order of nanoseconds at low temperature and tens of
 picoseconds at high
 temperature.\cite{Loren6,Loren7,Loren8,Loren9,Loren10,Loren11,three-four}
 In the ballistic spin-injection
configuration, based on the magnetic-field-induced spin relaxation channel arising from the anisotropy
of the $g$-factor of different L-valleys, Li {\em et al.}
have 
measured the SRT by means of the spin transport under a magnetic field in the order of
100 Gauss.\cite{Li-Ge4}
 The SRT was measured to be in the order of several hundreds
of nanoseconds at low temperature around 50~K.\cite{Li-Ge4} Moreover, sensitive electric field
dependence at weak electric fields ($\lesssim$ 50~V/cm) was observed,
 which was speculated to be the hot-electron
 effect.\cite{Li-Ge4} It is noted that the SRTs measured in the ballistic 
   spin-injection configuration are one to two orders of magnitude larger than thoses from
   the three- or
   four-terminal Hanle configuration, which are typically
     carried out with metallic doping concentrations (above the metal-to-insulator
     transition).
 With the electrodes attached to the surface of the
 sample in the three- or
   four-terminal Hanle configuration, it was speculated that the presence of interfaces and surface
 roughness could have much influence on 
the intrinsic spin relaxation.\cite{interface,Loren7,Loren8,
Loren9,Loren11,Loren10,interface6,interface7,surface} Besides, the presence
 of impurities in the tunnel barrier may also provide a possible explanation for the
 discrepancy.\cite{tunneling_im1,tunneling_im2,tunneling_im3,tunneling_im4,tunneling_im5}

 For the optical measurements,
 with the optical injection, several methods have
been developed and applied in Ge to determine the electron
 SRT.\cite{Guite,optical1,optical2,Pezzoli-Ge1,Pezzoli-energy,optical3}
 Guite {\em et al.} have developed a novel method
  based on the sensitively tuned radio-frequency coil to measure the temporal
  evolution of the magnetization
arising from the optical-injected spin polarization in Ge directly
 under the magnetic field (up to 80
Gauss).\cite{Guite} The temperature
dependence of the electron SRT was measured to decrease
from $\sim$5~ns to $\sim$2~ns from 100~K to 180~K.\cite{Guite} 
Furthermore, in the work of
  Lohrentz {\em et al.},\cite{optical3} by means of the resonant spin
  amplification method, the electron SRT exceeding 65~ns at 60~K was
  measured. Moreover, a peak for the SRT appeared in the
  temperature dependence, which was speculated to be the influence of
 the electron-impurity scattering. It is noted that the SRTs measured from the
    optical measurements\cite{Guite,optical3} are in the same order as those from the electric
    method in the spin-injection configuration.\cite{Li-Ge4}

Meanwhile, the theoretical studies for the electron spin dynamics in intrinsic Ge are in
  progress.\cite{Tang-Ge1,Li-Ge2,Li-Ge3} With the establishment of
 the electron-phonon interaction in the
L-valley by group-theory method, the SRT for the
 spin relaxation due to the electron-phonon
  scattering in the framework of the Elliott-Yafet (EY) mechanism\cite{Yafet,Elliott}
has been calculated in the non-degenerate limit.\cite{Tang-Ge1,Li-Ge2} It has been found
that the main spin relaxation source in intrinsic Ge comes from the inter-L valley electron-phonon
interaction.\cite{Tang-Ge1,Li-Ge2} However, up to now, there still exists
marked discrepancy between the
experimental observation and the theoretical calculation, especially at low
temperature at which the theoretical
calculations are two to three orders of magnitude 
larger than the experimental
observations.\cite{Guite,Li-Ge4,three-four,optical3} In the intrinsic sample, 
this was speculated to be the hot-electron effect in the
transport experiment.\cite{Li-Ge4} Therefore, a full investigation on 
 the hot-electron effect 
in the spin relaxation 
is needed. Moreover, although the hot-electron effect has been well studied in the charge transport
experiment in Ge,\cite{negative,hyperpure}
its influence on the spin relaxation has not yet been revealed. In the charge transport
experiment, it has been revealed that due to the weak electron-phonon interaction
in Ge, the hot-electron effect can be easily achieved especially at low
temperature.\cite{negative,hyperpure}
 This feature indicates that the electric field can influence the
behavior of spin relaxation easily due to the hot-electron effect.

In this work, we study the hot-electron effect in the spin relaxation in
intrinsic Ge
 by the kinetic
  spin Bloch equations (KSBEs)\cite{2001,jianhua23,jianhua15,wu-review} both analytically and
 numerically. When the electric field is weak, we compare our
calculations with the recent
transport experiment in the spin-injection configuration with weak electric
 field.\cite{Li-Ge4}
 Good agreements with the experimental data are obtained. It is
found that due to the weak electron-phonon interaction,\cite{negative,hyperpure}
 at low temperature, even small electric fields ($\lesssim$ 50~V/cm)\cite{Li-Ge4} can
cause obvious center-of-mass drift effect and hence significantly enhance the spin
relaxation, whereas the hot-electron effect is demonstrated to be less important. 
When the electric field is relatively strong ($0.5\lesssim E \lesssim 2$~kV/cm),
 we find that the SRT decreases with the increase
of the electric field because of the increase of the hot-electron temperature
and hence the enhancement of the electron-phonon scattering, whereas the drift
effect is shown to be marginal. Finally,
 with the intra-$\Gamma$ and intra-L-$\Gamma$ electron-phonon
interactions established by Liu {\em et al.},\cite{Liu} the influence of
the $\Gamma$ valley to the spin relaxation in the presence of the electric field
is revealed. We find that within the strength of the 
  electric fields we study, only a small fraction ($\lesssim
  5\%$) of the electron can be driven from the L valley
  to the 
  $\Gamma$ valley. Therefore, the influence of the electron spin dynamics in
  the $\Gamma$ valley to the whole system is marginal. We further reveal
    that in the intrinsic sample without impurity, 
the spin relaxation rates are
  the same for the $\Gamma$ and L valleys, and hence the spin dynamics in
  the L valley can be measured from the $\Gamma$ valley by the 
direct optical transition method.

This paper is organized as follows. In Sec.~{\ref{model}}, we set up the
  model and the KSBEs. In Sec.~{\ref{results}}, we present the main results
  obtained from the KSBEs both analytically and numerically. The calculated
  results under the weak electric field are compared with the experimental
 data (Sec.~{\ref{analytical}}).
 Then, under
  the relatively strong electric field,
 the influences of the hot-electron
  effect on the spin relaxation are presented (Sec.~{\ref{numerical}}).   
 We summarize in Sec.~{\ref{summary}}.
 
\section{Model and KSBE${\bold{\rm s}}$}
\label{model}
We start our investigation of the electron spin relaxation in 
 intrinsic Ge, where the four lowest valleys in
the conduction band are
located at the L points [$\frac{\pi}{a_0}(1,1,1)$, $\frac{\pi}{a_0}(-1,1,1)$,
$\frac{\pi}{a_0}(1,-1,1)$ and $\frac{\pi}{a_0}(1,1,-1)$ with
$a_0$ denoting the lattice constant]. The $\Gamma$ valley lies
energetically above the L valley with $\Delta E_{L}^{\Gamma}=0.151$~eV,\cite{Liu,Gamma} and the ${\rm X}$
valley lies further above the $\Gamma$ valley with $\Delta
E_{\Gamma}^{\rm X}=0.04$~eV.\cite{Pezzoli-energy}
In
the spherically symmetric approximation, the electron effective masses of the L
and $\Gamma$ valleys are $m_{L}^*=0.22 m_0$ ($m_l=1.588m_0$, $m_t=0.0815m_0$)\cite{Li-Ge2}
and $m_{\Gamma}^*=0.038 m_0$ ($m_l=m_t=0.038 m_0$),\cite{Gamma} respectively,
 with $m_0$ representing the free electron mass.
 In our investigation with the electric field $E \lesssim 2$~kV/cm,
 we do not consider the ${\rm X}$ valley as the fraction of the
 electron in these valleys are negligible. 

The KSBEs derived via the nonequilibrium Green
function method with the generalized Kadanoff-Baym
Ansatz read\cite{wu-review,2001,jianhua23,jianhua15,jianhua52}
\begin{equation}
\partial_t\rho_{\lambda {\bf k}_{\lambda}}=\partial_t\rho_{\lambda {\bf
    k}_{\lambda}}|_{\rm drift}+\partial_t\rho_{\lambda {\bf k}_{\lambda}}|_{\rm scat},
\end{equation}
in which $\rho_{\lambda{\bf k}_\lambda}$ is the density matrix of electrons with momentum
${\bf k}_\lambda$ in the $\lambda$ valley. Here, ${\bf k}_\lambda$ is defined in reference to
the valley center in the valley coordinate, whose $\hat{z}$-axis is
along the valley axis.\cite{Li-Ge2,Liu}
 The diagonal term $\rho_{\lambda{\bf
    k}_\lambda,\sigma\sigma}\equiv f_{\lambda{\bf
    k}_\lambda,\sigma}$ ($\sigma=\pm
1/2$) describes the
distribution of electrons in each spin band, and the off-diagonal term $\rho_{\lambda{\bf
    k}_{\lambda},\frac{1}{2}-\frac{1}{2}}=\rho_{\lambda{\bf
    k}_{\lambda},-\frac{1}{2}\frac{1}{2}}^\ast$ represents the coherence between the two spin bands.

In the KSBEs, the drift term is given by
\begin{equation}
\partial_t\rho_{\lambda {\bf
    k}_{\lambda}}|_{\rm drift}=-e\bf{E}\cdot\nabla_{\lambda {\bf k}_{\lambda}}\rho_{\lambda {\bf
    k}_{\lambda}},
\end{equation}
with $e< 0$.  $\partial_t\rho_{\lambda {\bf k}_{\lambda}}|_{\rm scat}$ stands for
the scattering term, which includes the electron-phonon (ep), electron-impurity (ei) and
electron-electron (ee) Coulomb scatterings:
\begin{equation}
\left.\partial_{t}\rho_{\lambda{\bf k}_\lambda}\right|_{{\rm scat}}=
\left.\partial_{t}\rho_{\lambda{\bf k}_\lambda}\right|_{{\rm ep}}+
\left.\partial_{t}\rho_{\lambda{\bf k}_\lambda}\right|_{{\rm ei}}+
\left.\partial_{t}\rho_{\lambda{\bf k}_\lambda}\right|_{{\rm ee}}.
\label{scat_all}
\end{equation}
Explicit forms of these scattering terms are shown in
Appendix~\ref{AA}.

The initial conditions at time $t=0$ are prepared as follows.
 We turn on the electric field at
  $t=-t_0$, where the system is in the equilibrium, and the density
  matrix are expressed as 
\begin{eqnarray}
&&f_{\lambda{\bf k}_{\lambda},\sigma}(-t_0)=\big\{\exp[(\varepsilon_{{\bf
    k}_{\lambda}}^{\lambda}-\mu)/(k_BT)]+1\big\}^{-1},\\
&&\rho_{\lambda{\bf k}_{\lambda},\frac{1}{2}-\frac{1}{2}}(-t_0)=\rho^*_{\lambda{\bf k}_{\lambda},-\frac{1}{2}\frac{1}{2}}(-t_0)=0,
\end{eqnarray}
with $\mu$ being the chemical potential for the electron at temperature $T$.
 The system is driven to the steady state before
$t=0$. Then at time $t=0$, the chemical potential is modified to obtain the
spin-polarized state with the spin polarization
$P_0=(N_{\uparrow}-N_{\downarrow})/(N_{\uparrow}+N_{\downarrow})$.
Here, $N_{\uparrow(\downarrow)}=\sum_{\lambda, {\bf k}_{\lambda}}f_{\lambda{\bf
     k}_{\lambda},\uparrow(\downarrow)}$ is 
  the electron density of spin-up (-down) state. 
 When $t\ge 0$,
the system relaxes from the spin-polarized state with the spin polarization
$P(t)=\sum_{\lambda {\bf k}_{\lambda}}\mbox{Tr}[\rho_{\lambda{\bf
    k}_{\lambda}}(t) {\sigma}_z]/n_e$ solved by the KSBEs.
  Here, $n_e=\sum_{\lambda,{\bf k}_{\lambda}}\mbox{Tr}[\rho_{\lambda{\bf
    k}_{\lambda}}(t)]$ is the electron density.

\section{Results} 
\label{results}

\subsection{Analytical results}
\label{analytical}
Before performing the full numerical calculation by solving the
KSBEs, we first investigate the spin relaxation analytically with the drift effect and
hot-electron effect explicitly included.
 It has been demonstrated that at relatively high temperature,
the dominant spin relaxation channel in Ge arises from the inter-L valley
electron-phonon interaction, in which the momentum dependencies of the matrix
elements for the electron-phonon interaction and the phonon energy are negligible.\cite{Li-Ge2} 

For the electron-phonon interaction, the matrix elements can be generally constructed in
this form,\cite{Tang-Ge1,Li-Ge2,Liu}
\begin{equation}
M^{\gamma}_{{\bf k}_{\lambda},{\bf k}'_{\lambda'}}
=A^{\gamma}_{{\bf k}_{\lambda},{\bf k}'_{\lambda'}}\hat{I}+{\bf B}^{\gamma}_{{\bf k}_{\lambda},{\bf
    k}'_{\lambda'}}{\cdot}{\bgreek \sigma},
\end{equation}
including both the spin-conserving and spin-flip parts for the interaction
  of electrons with 
the $\gamma$-branch phonon, where $\hat{I}$ and
${\bgreek \sigma}$ are $2\times 2$ unit and Pauli matrices.
 The SRT due to the inter-L valley electron-phonon interaction can be directly deduced from the
scattering term [Eq.~(\ref{scat_ep})] in the KSBEs ($\hbar \equiv 1$ throughout this paper)\cite{Li-Ge2,Tang-Ge1} 
\begin{eqnarray}
\nonumber
\frac{1}{\tau_s^{\rm ep}}&=&\frac{2\pi}{Vd}\sum_{{\bf k}_{\lambda}}\sum_{{\bf
    k}'_{\lambda'},\lambda\neq \lambda'}\sum_{{\gamma},\pm}\delta(\pm \Omega^{\gamma}_{{\bf k}'_{\lambda'}-{\bf
    k}_{\lambda}}+\varepsilon_{{\bf k}'_{\lambda'}}-\varepsilon_{{\bf
    k}_{\lambda}})\\
\nonumber
&&\mbox{}\times \frac{1}{\Omega^{\gamma}_{{\bf k}'_{\lambda'}-{\bf
      k}_{\lambda}}} N_{{\bf  
    k}'_{\lambda'}-{\bf k}_{\lambda}}^{{\gamma},\pm}(|B_{{\bf k}_{\lambda},{\bf
    k}'_{\lambda'}}^{{\gamma},x}|^2
+|B^{{\gamma},y}_{{\bf k}_{\lambda},{\bf k}'_{\lambda'}}|^2)\\
&&\mbox{}\times(f^d_{{\bf
    k}_{\lambda}\uparrow}-f^d_{{\bf k}_{\lambda}\downarrow})\Big[\sum_{{\bf k}_{\lambda}}(f^d_{{\bf
    k}_{\lambda}\uparrow}-f^d_{{\bf k}_{\lambda}\downarrow})\Big]^{-1},
\label{Fermi_Golden}
\end{eqnarray}
in which, $V$ and $d$ are the volume and density of the
  crystal, respectively; ${\gamma}$ labels the associated phonon branches in the ${\rm X}$
point connecting the two L valleys, including ${\rm X}_{1a}$, ${\rm X}_{1b}$, ${\rm X}_{4a}$ and ${\rm X}_{4b}$
phonons;\cite{Li-Ge2,Tang-Ge1,Liu}
 $N_{{{\bf  
    k}^{\prime}_{\lambda^{\prime}}-{\bf k}_\lambda}}^{\gamma,\pm}=N^{\gamma}_{{{\bf  
    k}^{\prime}_{\lambda^{\prime}}-{\bf k}_\lambda}}+\frac{1}{2}\pm
\frac{1}{2}$ and $N^{\gamma}_{{{\bf  
    k}^{\prime}_{\lambda^{\prime}}-{\bf k}_\lambda}}=\big\{\exp[\Omega^{\gamma}_{{\bf  
    k}^{\prime}_{\lambda^{\prime}}-{\bf k}_\lambda}/(k_B T)]-1\big\}^{-1}$ is the
Bose distribution of phonons with energy $\Omega^{\gamma}_{{\bf  
    k}^{\prime}_{\lambda^{\prime}}-{\bf k}_\lambda}$; $f^d_{{\bf
    k}_{\lambda}\sigma}$
 is the drifted Fermi distribution
function of the electron in the steady state, which reads as{\cite{Yang_hot,Lei_hot,hot4}
\begin{equation}
f_{{\bf k}_{\lambda}\sigma}^d=\Big\{\exp\big[\frac{({\bf
    k}_{\lambda}-m^*_{\lambda}{\bf v}_{\lambda})^2}
{2m^*_{\lambda}}-\mu_{\sigma}\big]/(k_BT_e)+1\Big\}^{-1},
\label{steady_state}
\end{equation} 
  with ${\bf v}_{\lambda}$ being the steady-state drift velocity of $\lambda$
  valley and $T_e$ standing for 
 the hot-electron temperature. One notices that the drifted Fermi distribution
function [Eq.~(\ref{steady_state})] is widely used in the treatment of nonlinear transport in
semiconductors.\cite{Yang_hot,Lei_hot,hot4} In Eq.~(\ref{steady_state}),
 ${\bf v}_{\lambda}$ is numerically obtained from the steady-state value
of ${\bf v}_{\lambda}(t)\equiv\sum_{{\bf k}_{\lambda}\sigma}[f_{{\bf
      k}_{\lambda},\sigma}(t)\hslash{{\bf k}_{\lambda}}/m_{\lambda}]/\sum_{{\bf k}_{\lambda}\sigma}f_{{\bf
      k}_{\lambda},\sigma}(t)$, and $T_{e}$ is
obtained by fitting the Boltzmann tail of the numerically calculated steady-state electron
distribution of each valley from the KSBEs.

When the electric field is weak with
 the condition $\frac{1}{2}m^{*}_L{\bf v}_{\lambda}^2\ll k_BT_e$ or $\frac{1}{2}m^{*}_L{\bf
    v}_{\lambda}^2\beta_e\ll 1$ [$\beta_e=1/(k_BT_e)$] satisfied, we expand the steady-state distribution
function to the order of ${\bf v}^2_{\lambda}$.  Accordingly, in the small spin polarization and
non-degenerate limit,  
 the SRT due to the inter-L valley electron-phonon interaction [Eq.~(\ref{Fermi_Golden})] can be obtained   
\begin{widetext}
\begin{equation}
\frac{1}{\tau_s^{\rm ep}}=\sum_{\gamma}\frac{1}{\tau_s^{\gamma}({\bf
    v}_{\lambda}=0)}\big[1+F(\beta_e\Omega_{\gamma})\beta_e
m^*_L{\bf v}_{\lambda}^2\big],
\label{non_scat_drift}
\end{equation}
with
\begin{equation}
\frac{1}{\tau_s^{\gamma}({\bf
    v}_{\lambda}=0)}=\frac{\sqrt{\beta_e}}{8d}\Big(\frac{2m^*_L}{\pi}\Big)^{3/2}
K_1\Big(\frac{\beta_e\Omega_{\gamma}}{2}\Big)
\big(|B^{\gamma,x}|^2+|B^{\gamma,y}|^2\big)\frac{A_{\gamma}}{\sinh(\beta\Omega_{\gamma}/2)}
\cosh\big(\frac{\beta-\beta_e}{2}\Omega_{\gamma}\big)
\label{non_scat}
\end{equation}
\end{widetext}
and 
\begin{equation}
F(x)=\frac{\displaystyle
   \sqrt{\pi}}{\displaystyle 2x}
\frac{\displaystyle U(-0.5,-2,x)}{\displaystyle K_1(x/2)}
\exp({-x/2})-\frac{\displaystyle 1}{\displaystyle 2}.
\label{factor}
\end{equation}

In Eqs.~(\ref{non_scat_drift}) and (\ref{non_scat}), for the inter-L valley
scattering with $|{\bf K}_{L_i}-{\bf K}_{L_j}|\gg |{\bf k}_{\lambda}|$ ($i\ne
j$),
 the momentum dependencies of the matrix
elements for the electron-phonon interaction and the phonon energy are
negligible,\cite{Li-Ge2} and hence their labels of momentum are omitted.
 $\beta=1/(k_BT)$ with $T$ being the lattice temperature. $\rm
A_{\gamma}=16$ ($\rm
A_{\gamma}=8$) for ${\rm X}_{1a}$ and ${\rm X}_{1b}$ (${\rm X}_{4a}$ and ${\rm
  X}_{4b}$) phonons.\cite{Li-Ge2} 
 $K_1(x/2)$ and
$U(-0.5,-2,x)$ are the modified Bessel function of the second kind and Tricomi's 
confluent hypergeometric function, respectively. 

It is noted that our results in Eqs.~(\ref{non_scat_drift}) and (\ref{non_scat})
 are ${3}/{8}$ times of that in the work of Li
{\em et al.} [Eq.~(10) in Ref.~\onlinecite{Li-Ge2}] when $\beta=\beta_e$ and
${\bf v}_{\lambda}=0$. However, by means of the same equation
  [Eq.~(\ref{Fermi_Golden})] here,
 the SRT due to the $f$-process of the electron-phonon scattering in
  Silicon [Eq.~(18) in Ref.~\onlinecite{Li-Si1}] can be recovered.
 Moreover, our analytical results are further confirmed by the numerical ones
as shown in the next section (refer to Fig.~\ref{figyw1}). Therefore, the results
in the work of Li
{\em et al.} [Eq.~(10) in Ref.~\onlinecite{Li-Ge2}]
    underestimate the spin relaxation rate in
    intrinsic Ge, and should be corrected.

 From the analytical results, it is seen that the spin relaxation rates
 are independent on the electron density in the non-degenerate regime.
 Moreover,
 the effect of the drift and hot-electron effects in the spin relaxation
under the weak electric field can be obtained.
 First of all, it can be shown that the drift effect can enhance the spin
 relaxation with the factor $F(x)$ [Eq.~(\ref{factor})] always larger than zero.
 However, under the weak electric field, with $F(x)$ decreasing from
$1/6$
monotonically to 0 when $x$ increases from 0, it shows that the influence of the
drift effect due to the electric field to the spin relaxation is marginal when
 $F(x)\beta_em^{*}_L{\bf
    v}_{\lambda}^2\ll 1$. In this situation, the electron SRT in the presence of the
  electric field can be approximately obtained from Eq.~(\ref{non_scat}),
  and hence the
  hot-electron effect has dominant influence on the spin relaxation. When the hot-electron temperature increases,
 $\sqrt{\beta_e}$ decreases slowly, 
 $K_1\big({\beta_e\Omega_{\gamma}}/{2}\big)$ increases rapidly and
 $\cosh\big[(\beta-\beta_e)\Omega_{\gamma}/2\big]$ increases slowly. Hence,
 the SRT decreases with the increase of the hot-electron temperature.
  It can also be obtained that when the
  lattice temperature increases,
 ${[\sinh(\beta\Omega_{\gamma}/2)]}^{-1}$ increases and
 $\cosh\big[{(\beta-\beta_e)}\Omega_{\gamma}/2\big]$ (when
 $\beta_e\lesssim\beta$) and $\sqrt{\beta_e}$ decrease slowly.
 Hence, the SRTs decreases with the increase of the
 lattice temperature.\cite{Li-Ge2}

\subsection{Numerical results}
\label{numerical}
In this section, we present our results obtained by numerically
solving the KSBEs following the scheme laid out in
Refs.~\onlinecite{Jianhua,Peng,Hua}. 
All parameters including the material parameters, band structure and phonon parameters used in our computation are
listed in Table I. During the calculation, the impurity density $n_i$ is set to
  be zero in the intrinsic sample. Furthermore, in the intrinsic sample, the electron can be injected by means of the
  electrical and optical methods,\cite{Guite,Li-Ge4,optical1,optical2,Pezzoli-Ge1,
Pezzoli-energy,optical3,Tang-Ge1,Li-Ge2,Li-Ge3} whose density $n_e$ varies from $10^{13}$ to $10^{17}$~cm$^{-3}$.
\begin{table}[htbp]
  \caption{Parameters used in the computation.}
  \label{material}
\begin{tabular}{ll|ll}\hline\hline
    $m_{L}^*/m_0$&\;\;\;\;$0.22^a$\;&\;\;$\Omega_{\rm L1}({\rm meV})$\;&\;\;\;\;\;$29.3^c$\\[4pt]
    $m_{\Gamma}^*/m_0$&\;\;\;\;$0.038^b$\;&\;\;$\Omega_{\rm L3}({\rm meV})$&\;\;\;\;\;$7.2^c$\\[4pt]
    $\Delta E_L^{\Gamma}({\rm eV})$\;\;&\;\;\;\;$0.151^{b,c}$&\;\;$\Omega_{\rm L2'}({\rm meV})$&\;\;\;\;\;$25.6^c$\\[4pt]
    $\Delta E_{\Gamma}^{X}({\rm eV})$\;\;&\;\;\;\;$0.04^d$&\;\;$\Omega_{\rm L3'}({\rm meV})$&\;\;\;\;\;$35.8^c$\\[4pt]
    $d(10^3{\rm kg/cm^3})$\;&\;\;\;\;$5.323^e$&\;\;$D_{\rm X1s}({\rm eV/nm})$&\;\;\;\;\;$0.18^c$\\[4pt]
    $\kappa_0$&\;\;\;\;$16.0^f$\;&\;\;$D_{\rm X4s}({\rm eV/nm})$&\;\;\;\;\;$0.66^c$ \\[4pt]
    $v_{\rm LA}({\rm m/s})$&\;\;\;\;$4900^{e}$\;&\;\;$D_{\rm x1m}({\rm eV/nm})$&\;\;\;\;\;$6.56^c$\\[4pt]
    $v_{\rm TA}({\rm m/s})$&\;\;\;\;$3500^{e}$\;&\;\;$A_{\rm L2'}({\rm eV/nm})$&\;\;\;\;\;$18.21^c$\\[4pt]
    $\Omega_{\rm LO,\Gamma}({\rm meV})$\;\;&\;\;\;\;$38.2^c$&\;\;$B_{L3'y}({\rm eV/nm})$&\;\;\;\;\;$-0.35\, i^c$\\[4pt]
    $\Omega_{\rm X1}({\rm meV})$\;&\;\;\;\;$28.4^c$&\;\;$P_0$&\;\;\;\;\;$30\%^d$\\[4pt]
    $\Omega_{\rm X4}({\rm meV})$\;&\;\;\;\;$33.3^c$ &\;\;$t_0$ (ps)&\;\;\;\;\;$30$\\[4pt]
    $\Omega_{\rm X3}({\rm meV})$\;&\;\;\;\;$10.2^c$ &\;\;&\;\;\;\;\;\;\;\\[4pt]
    \hline\hline
\end{tabular}\\
\hspace{-6.8cm}$^a$Ref.~\onlinecite{Li-Ge2}.\\
\hspace{-6.76cm}$^b$Ref.~\onlinecite{Gamma}.\\
\hspace{-6.76cm}$^c$Ref.~\onlinecite{Liu}.\\
\hspace{-6.72cm}$^d$Ref.~\onlinecite{Pezzoli-energy}.\\
\hspace{-6.7cm}$^e$Ref.~\onlinecite{Tang-Ge1}.\\
\hspace{-6.7cm}$^f$Ref.~\onlinecite{Landolt}.\\
\end{table}

Before the full numerical investigation, we first present the
  electric field dependence of the hot-electron temperature and steady-state
  drift velocity when the electron density $n_e=5\times 10^{17}$~cm$^{-3}$. We find that the spin relaxation
in intrinsic Ge in the
presence of the electric field can be divided into three regimes according to the hot-electron
temperature: (1) when $E\lesssim 1.4$~kV/cm, only the L valley is relevant for the
spin relaxation; (2) when $1.4 \lesssim E\lesssim 2$~kV/cm, the $\Gamma$ valley
becomes relevant; (3) when $E\gtrsim 2$~kV/cm, the X valley becomes
relevant. The boundaries between different regimes are shown in
Fig.~\ref{figyw1}, where the electric field dependence of the hot-electron temperature 
 for electrons in the L valley with $n_e=5\times 10^{17}$~cm$^{-3}$ at different
 temperatures
 ($T=$60, 150 and 300~K) is presented (the steady-state drift
velocity is also shown). 
 It can be seen that
both the hot-electron temperature and the steady-state drift
velocity increase with increasing the electric field.
 For electrons with the Boltzmann distribution
and marginal drift effect ($\frac{1}{2}m^{*}_L{\bf
    v}_{\lambda}^2\beta_e\ll 1$),
its average energy is estimated to be $\bar{E}=\frac{3}{2}k_BT_e$. Accordingly, the
hot-electron temperatures at which electrons can be driven to the $\Gamma$
and X valleys efficiently are estimated to be $\Delta E_L^{\Gamma}/(\frac{3}{2}k_B)\approx
1200$~K and $(\Delta E_L^{\Gamma}+\Delta E_{\Gamma}^{X})/(\frac{3}{2}k_B)\approx 1500$~K,
corresponding to $E\approx 1.4$~kV/cm (shown as the black vertical
dashed line) and $2$~kV/cm (shown as the pink vertical dashed line) in
Fig.~\ref{figyw1},
respectively.  We emphasize that this division of the system is irrelevant
  to the electron density in the non-degenerate regime. We further perform the
  calculation with $n_e=10^{13}$~cm$^{-3}$ with the Fermi
  temperature $T_F\approx 1$~K, which are not
  shown in Fig.~\ref{figyw1} for their coincidences with the ones with
  $n_e=5\times 10^{17}$~cm$^{-3}$ ($T_F\approx 45$~K).
 In this work, we study the spin relaxation in the presence of
the electric field $E$ up to 2~kV/cm under which the L and $\Gamma$ valleys are
relevant.

\begin{figure}[htb]
  {\includegraphics[width=8.6cm]{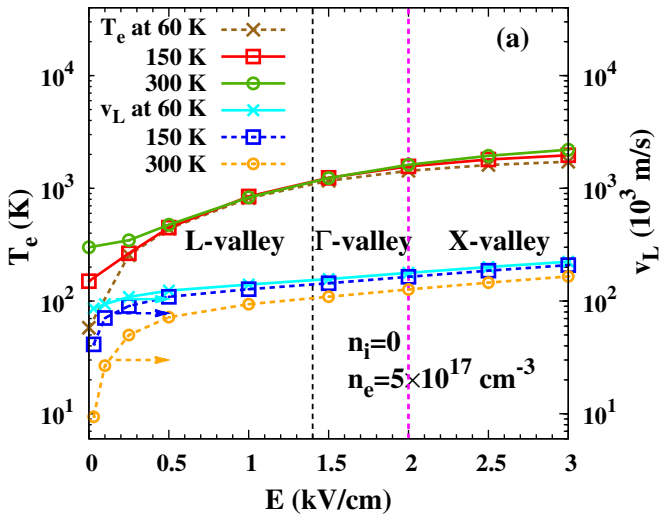}}
  \caption{(Color online) Electric field dependence of the 
    hot-electron temperature 
 and steady-state drift
velocity (note the scale of both curves
is on the right hand side of the frame) for electrons in the L valley
 at different temperatures (60, 150 and 300~K) with the electron density
 $n_e=5\times 10^{17}$~cm$^{-3}$. The pink (black) vertical
dashed line at $E\approx 1.4$~kV/cm ($E\approx 2$~kV/cm) corresponds to the boundary
at which the
$\Gamma$ (X) valley becomes relevant for the spin relaxation.}
\label{figyw1}
\end{figure}

In Fig.~\ref{figyw1}, some features should be further addressed. 
 It is shown that when $E\lesssim 0.5$~kV/cm, with the increase of the electric
field, the
hot-electron temperature increases slowly, whereas the drift velocity increases
rapidly (even when $E\lesssim 0.1$~kV/cm); when $0.5 \lesssim E\lesssim 3$~kV/cm,
the hot-electron temperature increases, whereas the drift velocity becomes
saturate. Specifically, when the electric field is extremely weak
  $E\lesssim 0.1$~kV/cm, the hot-electron temperature equals to the lattice one
  approximately
 at different temperatures (60, 150 and 300~K).
 These features are consistent with the experimental observations about the
drift velocity.\cite{negative,hyperpure} Accodingly, the electric field can 
be divided
into weak- ($E\lesssim 0.5$~kV/cm) and
strong-electric-field ($0.5 \lesssim E\lesssim 2$~kV/cm) regimes.

 \subsubsection{Spin relaxation under weak electric field: comparison with
   experiments}
\label{weak}

As mentioned in the introduction, the experiments on the electron spin relaxation in
intrinsic Ge 
have been carried out recently by several methods including
both the electrical\cite{Loren6,Loren7,Loren8,Loren9,Loren10,Loren11,three-four,Li-Ge4} and optical
ones.\cite{optical1,optical2,Pezzoli-Ge1,Pezzoli-energy,Guite,optical3}
Up to now, there still exists marked discrepancy between the experimental observations
and the theoretical
calculations in the intrinsic sample,
 especially at low temperature around 30~K.\cite{Guite,Li-Ge4,optical1,optical2,Pezzoli-Ge1,
Pezzoli-energy,optical3,Tang-Ge1,Li-Ge2,Li-Ge3} 
This motivates us
to carry out the full calculations
including the electron-phonon and electron-electron Coulomb 
scatterings in the presence of the electric field. In the works of Lohrentz {\em et
   al.}\cite{optical3} and Li {\em et al.},\cite{Li-Ge4} the electron density is
in the order of $n_e=10^{13}$~cm$^{-3}$.\cite{mobility} Accordingly, when comparing with the
experiments,\cite{Li-Ge4,optical3} we choose the typical electron density
 to be $n_e=10^{13}$~cm$^{-3}$ here, which lies in the non-degenerate regime.
 From Eq.~(\ref{non_scat_drift}), we emphasize that
   the chosen electron density does not influence the spin relaxation
   rates in the non-degenerate regime.\cite{Li-Ge2,wu-review}
Our calculation further shows that even
     the electron density is taken to be 
$n_e=5\times 10^{17}$~cm$^{-3}$ realized in the optical
experiment,\cite{Pezzoli-Ge1} which lies in the crossover region of the
non-degenerate and degenerate regimes at low
temperature with $T_F\approx 45$~K, the spin relaxation rates are marginally
influenced.
 The results in the presence of the weak
electric field ($E\lesssim 50$~V/cm) are summarized in
Figs.~\ref{figyw2}(a) and (b). 

\begin{figure}[htb]
  {\includegraphics[width=7.5cm]{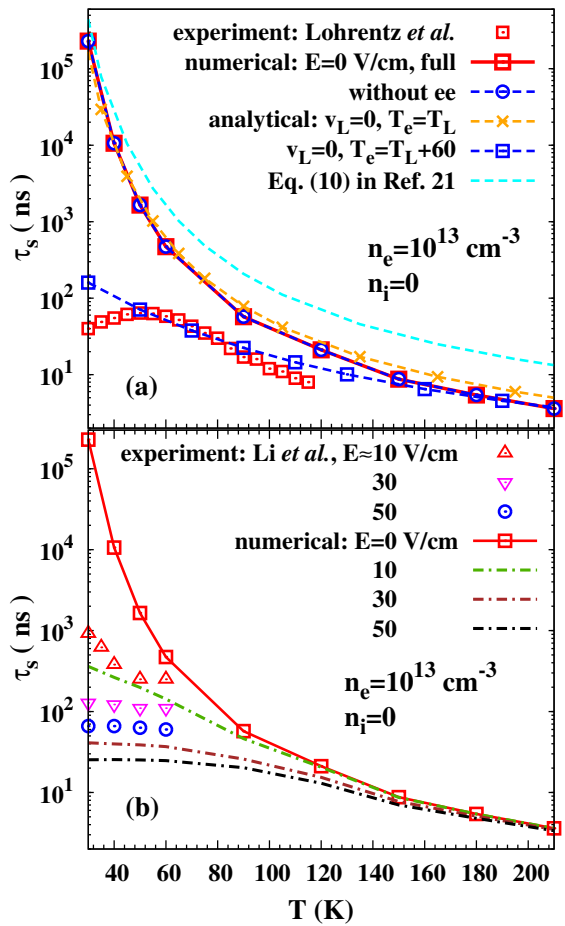}}
  \caption{(Color online) Electron SRTs in intrinsic Ge as function of
    temperature.
 (a), experimental results: red open squares correspond to the experiment of Lohrentz {\em et al.}
    (Ref.~\onlinecite{optical3}). Numerical results:
 the red solid curve with squares (blue
dashed curve with circles) corresponds to the numerical
result with all the scatterings (without the
electron-electron scattering).
 Analytical result: the orange dashed curve
 with crosses
 represents the analytical results calculated from Eq.~(\ref{non_scat}) by 
 setting $T_e=T$. The results
  calculated according to Eq.~(10) in Ref.~\onlinecite{Li-Ge2} 
by Li {\em et
    al.} are plotted by the
  cyan dotted curve using parameters in work of Liu {\em et
    al.} (simply 8/3 times of the orange
 dashed curve
 with crosses).\cite{Liu}
 (b), the experimental results of Li {\em et al.}
 (Ref.~\onlinecite{Li-Ge4}) are
    shown as red open upward triangles for $E\approx 10$~V/cm, pink open
    downward triangles
 for $E\approx 30$~V/cm, and blue open 
circles for $E\approx 50$~V/cm, respectively. Numerical results: when $n_i=0$, 
the green, brown and black dot-dashed curves represent
   the SRTs calculated numerically with $E\approx$10,
 30 and 50~V/cm, respectively.}
\label{figyw2}
\end{figure}

In Fig.~\ref{figyw2}(a),  when electric field $E=0$ and the impurity density
$n_i=0$,
 it can be seen that the SRTs with (the red solid curve with
squares) and without (the blue dashed curve with circles) the
electron-electron
scattering coincide with each other. Therefore, the influence
 of the electron-electron scattering to the spin
 relaxation in the L valley is marginal. Moreover, the numerical results with
 all the scatterings (shown by the
red solid curve with squares) 
are consistent with the analytical results calculated from Eq.~(\ref{non_scat}) by 
 setting $T_e$ equal to the lattice temperature $T_L$ (the orange dashed curve with crosses),
 in which only the inter-L valley scattering is included. This shows that the inter-L valley
 scattering is dominant for the spin relaxation.\cite{Li-Ge4} The result
calculated according to Eq.~(10) using the parameters in the work of Liu
  {\em et al.}\cite{Liu}
 in Ref.~\onlinecite{Li-Ge2} 
by Li {\em et al.} is also plotted by the
  cyan dotted curve, showing marked discrepancy compared with the full 
  numerical result.  Furthermore, it can be seen that the
 SRT decreases monotonically with the increase of the temperature.
 This is because with the
increase of the temperature, the electron-phonon scattering is enhanced, which
enhances
 the spin relaxation due to the EY mechanism.\cite{Yafet,Elliott}
 
In Fig.~\ref{figyw2}(a), when the electric field is zero, we compare the
temperature
 dependence of the electron SRTs
obtained from our model with the optical experiments by Lohrentz {\em et
   al.} (red open squares).\cite{optical3} One notices that these optical observations
deviate orders of magnitude from the calculations at $E$=0.
 However, in
Fig.~\ref{figyw2}(a), we show that when using an effective hot-electron
temperature $T_e=(T_L+60)$ K, the analytical results
(green dashed curve with squares) are comparable with the
experimental ones especially at $T\gtrsim 50$ K. 
Accordingly, this provides a possible explanation for the enhancement of the spin relaxation in the optical
  experiment.\cite{optical3} In the
 optical
 experiments,
 the electrons are first optically injected to the $\Gamma$ valley, which are then
scattered to the L and X valleys through the inter-valley electron-phonon
scattering.\cite{optical1,optical2,Pezzoli-Ge1,Pezzoli-energy,Guite,optical3}
 Accordingly, with the weak intra-valley electron-phonon
 scattering,\cite{negative,hyperpure,Liu} 
 the cooling process is suppressed, and hence the hot-electron
effect can arise easily.\cite{Pezzoli-energy} Therefore, the enhancement of the spin relaxation in
the optical experiments may arise from the hot-electron effect, which has been
discussed in GaAs.\cite{Jianhua,Ka,Ka_explain}

When a weak electric field ($\lesssim$ 50~V/cm) is applied, in Fig.~\ref{figyw2}(b),  
 we further compare our calculations with Li {\em et al.}
 (red open upward triangles for $E\approx 10$ V/cm, pink open downward triangles
 for 30 V/cm, and blue open circles for 50~V/cm).\cite{Li-Ge4}
  In the experiment of
 Li {\em et al.},\cite{Li-Ge4} the samples are
 nominally undoped. Therefore, we perform the calculation with the impurity
   density $n_i=0$. It is
 shown that the
 numerical results (the green, brown and black dot-dashed curves represent
   the SRTs calculated numerically with $E\approx$10,
 30 and 50~V/cm, respectively) are comparable with the experimental values, with
 the theoretical results being $1/3$ to $1/2$ of the experimental ones.
 Moreover, we further find that with the weak electric field
 ($E\lesssim 50$~V/cm), the electron temperature equals to the lattice
 one approximately, i.e., $T_e\approx T$, whereas $\frac{1}{2}\beta_e
 m_L^*{\bf v}_L^2\approx 2.4$ at 30~K and $\frac{1}{2}\beta_e
 m_L^*{\bf v}_L^2\gtrsim 1$ between 40 and 60~K when $E\approx 50$~V/cm. This shows that the
 hot-electron effect for the system in the presence of the weak electric field at
 low temperature is marginal, whereas the drift effect is obvious.
 Therefore, this significant enhancement
 of the spin relaxation in the presence of the weak electric field at low
 temperature arises from the drift effect rather than the hot-electron effect. 

The enhancement of the spin relaxation due to the drift effect can be
  understood as follows. It has been addressed that the phonon-induced
 intervalley scattering between the L-valleys is the dominant spin-flip
 mechanism.\cite{Li-Ge4} At low temperature, only phonon
 emission is feasible
 (involved phonon energy is about 30 meV),\cite{Liu,Li-Ge2}
 which needs the electron in the initial state to be 30 meV or more above
 the conduction-band edge. The drift
 effect can shift the electron distribution and hence
 increase the population of electrons with energies
 more than 30~meV,  needed to ignite the intervalley process. Accordingly,
due to the drift effect, the inter-L valley spin-flip scattering is enhanced and
hence the spin relaxation.

\subsubsection{Hot-electron effect on the spin relaxation under relatively
  strong electric field}

In the presence of the 
relatively strong  
electric field ($0.5\lesssim E \lesssim 3$~kV/cm),
 we first study the electric field dependence of the electron spin relaxation in Ge
  at different temperatures $T=$150 and 400~K both numerically and analytically. In this subsection, in order to study
  the influence of the $\Gamma$ valley on the spin relaxation, the electron density is set to be
  $5\times 10^{17}$~cm$^{-3}$,\cite{Pezzoli-energy} with more electrons driven
  to the $\Gamma$ valley compared to the case with low electron density.}
 In Fig.~\ref{figyw3}, the SRTs for the electrons in the L valley with (the red solid curves) and
without the $\Gamma$ valley (the dashed blue curves) are plotted against the
electric field, which is extended to 3~kV/cm, in the impurity-free situation.

\begin{figure}[htb]
  {\includegraphics[width=8.5cm]{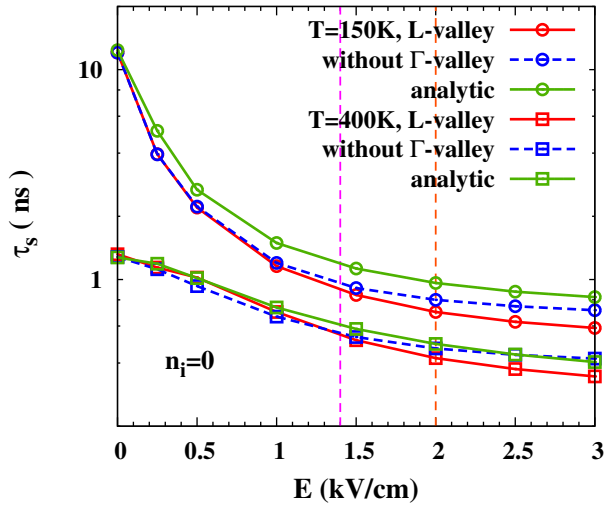}}
  \caption{(Color online) Electric field dependence of the electron SRTs in the
    L and $\Gamma$ valleys in the impurity-free situation at 150 and 400~K,
    respectively. The analytical results, obtained from
    Eqs.~(\ref{non_scat_drift}), (\ref{non_scat}) and (\ref{factor}) 
with the hot-electron temperature and the steady-state drift velocity
numerically calculated by the KSBEs, are shown by the green solid curve.}
\label{figyw3}
\end{figure}

 In Fig.~\ref{figyw3}, it can be seen that for both temperatures $T=150$ and
 400~K, the
SRTs in the L valley (the red solid curves) decrease monotonically
 with the increase of the electric field.
 It is noted that in Fig.~\ref{figyw1},
 we have presented the electric field dependencies of the hot-electron
temperature
 and the steady-state drift velocity, 
which increase with increasing electric field. From the analytical results in
Sec.~{\ref{analytical}}, both the
drift and hot-electron effects can enhance the spin relaxation. Their
contribution to the spin relaxation can be distinguished by
 using Eqs.~(\ref{non_scat_drift}), (\ref{non_scat}) and (\ref{factor}).
 With the
numerical values of the hot-electron temperature and the steady-state drift
velocity (Fig.~\ref{figyw1}),  
 the analytical results with (the green solid curves) and without
 (by setting ${\bf v}_{\lambda}=0$) the drift effect are
 calculated. We find that the two curves coincide
with each other at both 150 and 400 K, which are both consistent with the numerical ones. 
Therefore, the influence of the drift effect to the spin
relaxation at high temperature ($T\gtrsim 150$~K) is marginal.
 One concludes that the enhancement of the spin
relaxation mainly originates from the hot-electron effect, where the increase of the
hot-electron temperature enhances the electron-phonon interaction.  
 
 It has also been shown in Fig.~\ref{figyw1} that at relatively large electric field, the
   $\Gamma$ ($E\gtrsim 1.4$~kV/cm) and X valleys ($E\gtrsim 2$~kV/cm)
   become relevant. Therefore, in our study, we include the $\Gamma$ valley
   explicitly, and reveal its influence to the spin
   relaxation. It is shown
 in Fig.~\ref{figyw3} that at both 150 and 400~K,
 the SRTs in the L valley with (the red solid curves) and without
 (the blue dashed curves) the $\Gamma$
 valley are almost the same, especially at low electric field. This is because
 in the range of the electric field, the fractions of the electrons driven 
from the L to $\Gamma$ valley are small ($\lesssim 5\%$), and hence the
 $\Gamma$ valley plays a marginal role in 
 the spin relaxation of the whole system. Moreover, the numerical results show that
 the SRTs in the L and
$\Gamma$ (not shown in Fig.~\ref{figyw3})
 valleys are almost the same. This arises from the strong
inter-L-$\Gamma$ valley electron-phonon interaction and hence the frequent
exchange of electrons between the L and $\Gamma$ valleys.\cite{Peng,Hua} 
Accordingly, the spin dynamics in the L valley in the presence of the electric field in the
impurity-free situation can be measured from the $\Gamma$ valley
 by the standard optical
 method.\cite{Faraday_Kerr,Awschalom,Aws_rotation2,Aws_rotation3,Korn} It
   is noted that notwithstanding the fact that the calculated electric field is extended
   up to 3~kV/cm where the X
valley becomes relevant, luckily the population of
the electrons in that valley is still negligible.
 Therefore, the X valley has marginal influence on the spin relaxation in the
  L valley we calculate above. One expects that with the electric field
    further increasing, the spin relaxation may be
 significantly enhanced at the X valleys once the electrons populate
 the spin hot spots (where the spin mixing is large).\cite{hot_spot1,hot_spot2}

\section{Summary}
\label{summary}
In summary, we have investigated the hot-electron effect in the spin relaxation
of electrically injected electrons in
intrinsic Ge  
 by the KSBEs both analytically and
 numerically.\cite{wu-review} We first compare our
calculations with the recent
transport experiment by Li {\em et al.}\cite{Li-Ge4}
 in the spin-injection configuration when the electric field
is weak ($\lesssim$ 50 V/cm). 
 Our calculations agree with the experimental data fairly well, and hence can explain the
 marked discrepancy between the experiment of Li {\em et al.}\cite{Li-Ge4} and the
 previous theoretical calculations.\cite{Guite,Li-Ge4,three-four,optical3} It is 
revealed that at low temperature, even small electric fields ($\lesssim$ 50 V/cm)\cite{Li-Ge4} can
cause obvious center-of-mass drift effect due to the weak electron-phonon
interaction in Ge.\cite{negative,hyperpure} This can significantly enhance the spin
relaxation, whereas the hot-electron effect is demonstrated to be less important.

We then study the spin relaxation when the electric field is relatively strong ($0.5\lesssim E \lesssim 2$
kV/cm), under which the $\Gamma$ valley becomes relevant.  The electric
field dependence of the spin
relaxation is studied.
 We find that the SRT decreases with the increase
of the electric field. This is because with the increase of the electric field,
 the hot-electron temperature increases, and hence the electron-phonon
 scattering is enhanced.  Therefore, in the presence of the relatively strong
 electric field, the hot-electron effect has marked influence on the spin
 relaxation, whereas the drift
effect is shown to be marginal. 

 Finally, we further study the influence of
the $\Gamma$ valley on the spin relaxation in the presence of the electric
field.
 We find that within the strength of the 
  electric fields we study ($E\lesssim 2$~kV/cm), only a small fraction ($\lesssim
  5\%$) of the electron can be driven from the L valley
  to the 
  $\Gamma$ valley. Therefore, the influence of the electron spin dynamics in
  the $\Gamma$ valley to the whole system is marginal. Nevertheless, we
    find 
    that in the impurity-free situation, 
the spin relaxation rates are
  the same for the $\Gamma$ and L valleys, and hence the spin dynamics in
  the L valley can be measured from the $\Gamma$ valley by the standard optical methods.

\begin{acknowledgments}
This work was supported
 by the National Natural Science Foundation of China under Grant
No. 11334014 and  61411136001, the National Basic Research Program
 of China under Grant No.
2012CB922002 and the Strategic Priority Research Program 
of the Chinese Academy of Sciences under Grant
No. XDB01000000. One of the
authors (TY) would like to thank M. Q. Weng for suggestion on
analytical derivation.
\end{acknowledgments}

\begin{appendix}
\section{Scattering terms of the KSBEs}
\label{AA}
The scattering terms for the electron-phonon, electron-impurity and
electron-electron scatterings are shown as, 
\begin{eqnarray}
\nonumber
&&\left.\partial_{t}\rho_{\lambda{\bf k}_\lambda}\right|_{{\rm ep}}=-\pi\sum_{\lambda^\prime,{\bf
    k}^\prime_{\lambda^{\prime}},\pm}\sum_{\gamma}\delta(\pm\Omega^{\gamma}_{{\bf  
    k}^{\prime}_{\lambda^{\prime}}-{\bf k}_\lambda}+\varepsilon^{\lambda^\prime}_{{\bf  
    k}^{\prime}_{\lambda^{\prime}}}-\varepsilon^\lambda_{{\bf k}_\lambda}) \\ 
&&\hspace{0.2cm}\mbox{} \times
\nonumber
  \Big(N_{{\bf  
    k}^{\prime}_{\lambda^{\prime}}-{\bf k}_\lambda}^{\gamma,\pm} M^{\gamma}_{{\bf k}_\lambda,{\bf
    k}^\prime_{\lambda^{\prime}}}\rho^{>}_{\lambda^\prime{\bf k}^{\prime}_{\lambda^{\prime}}}M^{\gamma}_{{\bf
    k}^\prime_{\lambda^{\prime}},{\bf k}_\lambda}
    \rho^{<}_{\lambda{\bf k}_\lambda}-N_{{{\bf  
    k}^{\prime}_{\lambda^{\prime}}-{\bf k}_\lambda}}^{\gamma,\mp}\\
&&\hspace{0.2cm}\mbox{}\times M^{\gamma}_{{\bf k}_\lambda,{\bf
    k}^\prime_{\lambda^{\prime}}}\rho^{<}_{\lambda^\prime{\bf
        k}^{\prime}_{\lambda^{\prime}}}M^{\gamma}_{{\bf
    k}^\prime_{\lambda^{\prime}},{\bf k}_\lambda} 
    \rho^{>}_{\lambda{\bf k}_\lambda} \Big)+{\rm H.c.};
\label{scat_ep}
\end{eqnarray}
\begin{eqnarray}
\nonumber
&&\partial_{t}\rho_{\lambda{\bf k}_\lambda}|_{{\rm ei}}=\\
\nonumber
&&\mbox{}-\pi n_i Z_i^2\sum_{{\bf k}_\lambda^{\prime}}V_{{\bf k}_\lambda-{\bf
    k}_\lambda^{\prime}}^2(\hat{\Lambda}_{\lambda{\bf k}_\lambda,\lambda{\bf
    k}_\lambda^{\prime}}\rho^{>}_{\lambda{\bf
    k}_\lambda^{\prime}}\hat{\Lambda}_{\lambda{\bf k}_\lambda^{\prime},\lambda{\bf
    k}_\lambda}\rho^{<}_{\lambda{\bf k}_\lambda}  \\ 
\nonumber
&&\mbox{}-\hat{\Lambda}_{\lambda{\bf k}_\lambda,\lambda{\bf
    k}_\lambda^{\prime}}\rho^{<}_{\lambda{\bf
    k}_\lambda^{\prime}}\hat{\Lambda}_{\lambda{\bf k}_\lambda^{\prime},\lambda{\bf
    k}_\lambda}\rho^{>}_{\lambda{\bf k}_\lambda})\delta(\varepsilon^{\lambda}_{{\bf 
    k}_\lambda^{\prime}}-\varepsilon^\lambda_{{\bf k}_\lambda})+{\rm H.c.};\\
\label{scat_ei}
\end{eqnarray}
\begin{widetext}
\begin{eqnarray}
\nonumber
&&\left.\partial_{t}\rho_{\lambda{\bf k}_\lambda}\right|_{{\rm ee}}=-\pi \sum_{\lambda^\prime,{\bf k}_\lambda^{\prime},{\bf
    k}_{\lambda^{\prime}}^{\prime\prime}}V_{{\bf k}_\lambda-{\bf
    k}_\lambda^{\prime}}^2\delta(\varepsilon^{\lambda}_{{\bf
    k}_\lambda^{\prime}}-\varepsilon^\lambda_{{\bf
    k}_\lambda}+\varepsilon^{\lambda^\prime}_{{\bf
    k}_{\lambda^{\prime}}^{\prime\prime}}-\varepsilon^{\lambda^\prime}_{{\bf 
    k}_{\lambda^{\prime}}^{\prime\prime}-{\bf k}_\lambda+{\bf
    k}_\lambda^{\prime}}) \\
\nonumber
&& \mbox{}\times\Big\{\hat{\Lambda}_{\lambda{\bf k}_\lambda,\lambda{\bf
    k}_\lambda^{\prime}}\rho^{>}_{\lambda{\bf
    k}_\lambda^{\prime}}\hat{\Lambda}_{\lambda{\bf k}_\lambda^{\prime},\lambda{\bf
    k}_\lambda}\rho^{<}_{\lambda{\bf k}_\lambda}{\rm Tr}\big[\hat{\Lambda}_{\lambda^{\prime}{\bf k}_{\lambda^{\prime}}^{\prime\prime},\lambda^\prime({\bf
      k}_{\lambda^{\prime}}^{\prime\prime}-{\bf 
      k}_\lambda+{\bf k}_\lambda^{\prime})}\rho^{<}_{\lambda^\prime({\bf
      k}_{\lambda^{\prime}}^{\prime\prime}-{\bf 
      k}_\lambda+{\bf k}_\lambda^{\prime})}\hat{\Lambda}_{\lambda^\prime({\bf
      k}_{\lambda^{\prime}}^{\prime\prime}-{\bf 
      k}_\lambda+{\bf k}_\lambda^{\prime}),\lambda^\prime{\bf
      k}_{\lambda^{\prime}}^{\prime\prime}}\rho^{>}_{\lambda^\prime{\bf
      k}_{\lambda^{\prime}}^{\prime\prime}}\big]\\
&&\mbox{}-\hat{\Lambda}_{\lambda{\bf k}_\lambda,\lambda{\bf
    k}_\lambda^{\prime}}\rho^{<}_{\lambda{\bf
    k}_\lambda^{\prime}}\hat{\Lambda}_{\lambda{\bf k}_\lambda^{\prime},\lambda{\bf
    k}_\lambda}\rho^{>}_{\lambda{\bf k}_\lambda}{\rm Tr}\big[\hat{\Lambda}_{\lambda^{\prime}{\bf k}_{\lambda^{\prime}}^{\prime\prime},\lambda^\prime({\bf
      k}_{\lambda^{\prime}}^{\prime\prime}-{\bf 
      k}_\lambda+{\bf k}_\lambda^{\prime})}\rho^{>}_{\lambda^\prime({\bf
      k}_{\lambda^{\prime}}^{\prime\prime}-{\bf 
      k}_\lambda+{\bf k}_\lambda^{\prime})}\hat{\Lambda}_{\lambda^\prime({\bf
      k}_{\lambda^{\prime}}^{\prime\prime}-{\bf 
      k}_\lambda+{\bf k}_\lambda^{\prime}),\lambda^\prime{\bf
      k}_{\lambda^{\prime}}^{\prime\prime}}\rho^{<}_{\lambda^\prime{\bf
      k}_{\lambda^{\prime}}^{\prime\prime}}\big] \Big\}+ {\rm H.c.}. \ \ \ 
\label{scat_ee}
\end{eqnarray}
\end{widetext}
Here, $\rho_{\bf k}^{<}=\rho_{\bf k}$ and $\rho_{\bf
  k}^{>}=1-\rho_{\bf
  k}$. $\varepsilon^{L_i}_{{\bf k}_{L_i}}=k_{L_i}^2/(2m_{L}^*)$ and
  $\varepsilon^\Gamma_{{\bf k}_\Gamma}=k_\Gamma^2/(2m_\Gamma^*)+\Delta E_{L}^{\Gamma}$.
 In Eq.~(\ref{scat_ep}),
 for the intra-$\Gamma$
 valley electron-phonon scattering ($\lambda$=$\lambda^{\prime}$=$\Gamma$),
 the phonon branches include
$\rm TA_1$, $\rm TA_2$, $\rm LA$, $\rm
  TO_1$, $\rm TO_2$, and $\rm LO$ phonons, where the electron-phonon scattering
 with the ${\bf k}^0$-order is
  forbidden [for the spin-conserving (spin-flip) scattering, the matrix elements are
  proportional to ${\bf k}$ (${\bf k}^3$)].\cite{Liu}
 For the intra-L valley scattering ($\lambda$=$\lambda^{\prime}$=L),
 the phonon branches include $\rm
TO_1$, $\rm TO_2$, and $\rm LO$ phonons, and especially for the spin-flip scattering, the
matrix elements are proportional to ${\bf k}^3$, which are neglected in our study when the
inter-L valley scattering is dominant for spin relaxation.\cite{Liu,Li-Ge2}
 For the inter-$\Gamma$-L valley scattering, the phonon
 branches include $\rm L_1$, $\rm L_3$, $\rm L_{2'}$, and $\rm L_{3'}$
 phonons, where the electron-phonon interactions with 
${\bf k}^0$-order
 exist for both the spin-conserving and spin-flip scatterings.\cite{Liu}
 For the inter-L valley electron-phonon scattering, 
the phonon branches include $\rm X_1$ and $\rm X_4$ phonons, where the
electron-phonon interactions with ${\bf k}^0$-order also
 exist for both the spin-conserving and spin-flip scatterings,\cite{Liu,Li-Ge2} and especially we
 add the electron-phonon scattering with the 
${\bf k}$-order.\cite{Liu}

 In Eqs.~(\ref{scat_ei}) and (\ref{scat_ee}), 
 $n_i$ is the impurity density and $Z_i=1$ is the charge number of the
impurity; $V_{\bf{q}}$ is the screened Coulomb potential
under the random phase approximation.\cite{Jianhua,Peng,Hua,wu-review}
 The spin mixing $\hat{\Lambda}_{\lambda{\bf k}_\lambda,\lambda^{\prime}{\bf
    k}_{\lambda^{\prime}}^{\prime}}=\hat{I}
-\frac{1}{2}\big[S^{(1)}_{\lambda{\bf k}_\lambda}S^{(1)\dagger}_{\lambda{\bf
    k}_\lambda}
-2S^{(1)}_{\lambda{\bf k}_\lambda}S^{(1)\dagger}_{\lambda^{\prime}{\bf
    k}_{\lambda^{\prime}}^{\prime}}+S^{(1)}_{\lambda^{\prime}{\bf
    k}_{\lambda^{\prime}}^{\prime}}S^{(1)\dagger}_{\lambda^{\prime}{\bf
    k}_{\lambda^{\prime}}^{\prime}}\big]$. The matrix $S^{(1)}_{\lambda^{\prime}{\bf
    k}_{\lambda^{\prime}}^{\prime}}$ for the $\Gamma$ and L valleys can be found
in the $14\times14$ ${\bf k}\cdot{\bf p}$
Hamiltonian\cite{Gamma} and $16\times16$ one,\cite{Liu} whose explicit forms are
shown as follows. For the $\Gamma$ valley,
\begin{widetext} 
\begin{eqnarray} 
  S_{\Gamma{\bf k}_{\Gamma}}^{(1)}=
\left(\begin{array}{cccccc}  
\frac{\displaystyle 1}{\displaystyle \sqrt{2}}\frac{\displaystyle P^{+}}{\displaystyle E_g}&-\sqrt{\frac{\displaystyle 2}{\displaystyle 3}}\frac{\displaystyle P^{z}}{\displaystyle E_g}&-\frac{\displaystyle 1}{\displaystyle \sqrt{6}}\frac{\displaystyle P^{-}}{\displaystyle E_g}&0&-\frac{\displaystyle 1}{\displaystyle \sqrt{3}}\frac{\displaystyle P^{z}}{\displaystyle E_g+\Delta}&-\frac{\displaystyle 1}{\displaystyle \sqrt{3}}\frac{\displaystyle P^{-}}{\displaystyle E_g+\Delta} \\ 
0&\frac{\displaystyle 1}{\displaystyle \sqrt{6}}\frac{\displaystyle P^{+}}{\displaystyle E_g}&-\sqrt\frac{\displaystyle 2}{\displaystyle {3}}\frac{\displaystyle P^{z}}{\displaystyle E_g}&-\frac{\displaystyle 1}{\displaystyle \sqrt{2}}\frac{\displaystyle P^{-}}{\displaystyle E_g}&-\frac{\displaystyle 1}{\displaystyle \sqrt{3}}\frac{\displaystyle P^{+}}{\displaystyle E_g+\Delta}&\frac{\displaystyle 1}{\displaystyle \sqrt{3}}\frac{\displaystyle P^{z}}{\displaystyle E_g+\Delta} 
\end{array} \right),
\label{S_Gamma}
\end{eqnarray}
\end{widetext}
where $P^z=Pk_z$, $P^{\pm}=Pk_{\pm}=P(k_x\pm k_y)$ with $E_P=(2/m_0)P^2=26.3$
eV, $E_g=0.898$ eV and 
$\Delta=0.297$ eV.\cite{Gamma} 
For the L valley,
\begin{equation}
S_{L{\bf k}_{L}}^{(1)}=\Big(S_{L{\bf k}_{L}}^{\rm left},S_{L{\bf k}_{L}}^{\rm
  right}\Big),
\label{S_L}
\end{equation}
where
\begin{widetext} 
\begin{eqnarray} 
S_{L{\bf k}_{L}}^{\rm left}=\left(\begin{array}{ccccccc}
\frac{\displaystyle \alpha_4k_{-}}{\displaystyle
  E_{c1}-E_{c6}}&-\frac{\displaystyle P_4k_z}{\displaystyle
  E_{c1}-E_{c6}}&\frac{\displaystyle \sqrt{2}(P_3-\alpha_3)k_{-}}{\displaystyle
  E_{c1}-E_{c5}}&\frac{\displaystyle 2\sqrt{2}\alpha_3k_z}{\displaystyle
  E_{c1}-E_{c5}}&-\frac{\displaystyle (P_3+\alpha_3)k_{+}}{\displaystyle
  E_{c1}-E_{c4}}&\frac{\displaystyle (P_3+\alpha_3)k_{+}}{\displaystyle E_{c1}-E_{c4}}&\\
-\frac{\displaystyle P_4k_z}{\displaystyle E_{c1}-E_{c6}}&-\frac{\displaystyle
  \alpha_4k_{+}}{\displaystyle E_{c1}-E_{c6}}&\frac{\displaystyle
  2\sqrt{2}\alpha_3k_z}{\displaystyle E_{c1}-E_{c5}}&\frac{\displaystyle
  \sqrt{2}(\alpha_3-P_3)k_{+}}{\displaystyle E_{c1}-E_{c5}}&-\frac{\displaystyle
  (P_3+\alpha_{3})k_{-}}{\displaystyle E_{c1}-E_{c4}}&-\frac{\displaystyle
  (P_3+\alpha_{3})k_{-}}{\displaystyle E_{c1}-E_{c4}}
\end{array} \right),
\label{S_left}
\end{eqnarray}
and
\begin{eqnarray} 
S_{L{\bf k}_{L}}^{\rm right}=\left(\begin{array}{cccccc}
-\frac{\displaystyle \Delta_2}{\displaystyle E_{c1}-E_{c2}}&0&-\frac{\displaystyle
  (P_1+\alpha_1)k_{+}}{\displaystyle E_{c1}-E_{v1}}&\frac{\displaystyle
  (P_1+\alpha_1)k_{+}}{\displaystyle E_{c1}-E_{v1}}&-\frac{\displaystyle
  2\sqrt{2}\alpha_1k_z}{\displaystyle E_{c1}-E_{v2}}&\frac{\displaystyle
  \sqrt{2}(2\alpha_1-P_1)k_{-}}{\displaystyle E_{c1}-E_{v2}}\\
0&-\frac{\displaystyle \Delta_2}{\displaystyle E_{c1}-E_{c2}}&\frac{\displaystyle
  (P_1+\alpha_1)k_{-}}{\displaystyle E_{c1}-E_{v1}}&\frac{\displaystyle
  (P_1+\alpha_1)k_{-}}{\displaystyle E_{c1}-E_{v1}}&\frac{\displaystyle
  \sqrt{2}(P_1-2\alpha_1)k_{+}}{\displaystyle E_{c1}-E_{v2}}&-\frac{\displaystyle
  2\sqrt{2}\alpha_1k_z}{\displaystyle E_{c1}-E_{v2}}
\end{array} \right).
\label{S_right}
\end{eqnarray}
\end{widetext}
All the parameters in Eqs.~(\ref{S_left}) and (\ref{S_right}) are listed in Table
I of Ref.~{\onlinecite{Liu}}.

\end{appendix}


\begin{thebibliography}{0}
\bibitem{Loren6} C. Shen, T. Trypiniotis, K. Y. Lee, S. N. Holmes, R. Mansell, M.
Husain, V. Shah, X. V. Li, H. Kurebayashi, I. Farrer, C. H. de Groot,
D. R. Leadley, G. Bell, E. H. C. Parker, T. Whall, D. A. Ritchie,
and C. H. W. Barnes, Appl. Phys. Lett. {\bf 97}, 162104 (2010).
\bibitem{Loren7} H. Saito, S. Watanabe, Y. Mineno, S. Sharma, R. Jansen, S. Yuasa,
and K. Ando, Solid State Commun. {\bf 151}, 1159 (2011).
\bibitem{Loren8} Y. Zhou, W. Han, L. T. Chang, F. Xiu, M. Wang, M. Oehme, I. A.
Fischer, J. Schulze, R. K. Kawakami, and K. L. Wang, Phys. Rev.
B {\bf 84}, 125323 (2011).
\bibitem{Loren9} K. R. Jeon, B. C. Min, Y. H. Jo, H. S. Lee, I. J. Shin, C. Y. Park,
S. Y. Park, and S. C. Shin, Phys. Rev. B {\bf 84}, 165315 (2011).
\bibitem{Loren11} A. Jain, L. Louahadj, J. Peiro, J. C. Le Breton, C. Vergnaud,
A. Barski, C. Beign\'e, L. Notin, A. Marty, V. Baltz, S. Auffret,
E. Augendre, H. Jaffr\'es, J. M. George, and M. Jamet, Appl. Phys. Lett. {\bf 99}, 162102 (2011).
\bibitem{Loren10} K. Kasahara, Y. Baba, K. Yamane, Y. Ando, S. Yamada, Y. Hoshi,
  K. Sawano, M. Miyao and K. Hamaya, J. Appl. Phys. {\bf 111}, 07C503 (2012).
\bibitem{three-four} L. T. Chang, W. Han, Y. Zhou, J. Tang, I. A. Fischer, M. Oehme,
J. Schulze, R. K. Kawakami and K. L. Wang, Semicond. Sci. Technol. {\bf 28} 015018 (2013).
\bibitem{Li-Ge4} P. Li, J. Li, L. Qing, H. Dery, and I. Appelbaum, Phys. Rev. Lett. {\bf 111}, 257204
  (2013).

\bibitem{Guite} C. Guite and V. Venkataraman, Phys. Rev. Lett. {\bf 107}, 166603
  (2011); Appl. Phys. Lett. {\bf 101}, 252404 (2012).
\bibitem {optical1} C. Hautmann, B. Surrer, and M. Betz, Phys. Rev. B {\bf
    83}, 161203 (R) (2011).  
\bibitem{optical2} C. Hautmann and M. Betz,  Phys. Rev. B {\bf
    85}, 121203 (R) (2012).
\bibitem{Pezzoli-Ge1} F. Pezzoli, F. Bottegoni, D. Trivedi, F. Ciccacci,
  A. Giorgioni, P. Li, S. Cecchi, E. Grilli, Y. Song, M. Guzzi,  {H. Dery}, and
  G. Isella, Phys. Rev. Lett. {\bf 108}, 156603 (2012).
\bibitem{Pezzoli-energy} F. Pezzoli, L. Qing, A. Giorgioni, G. Isella,
  E. Grilli, M. Guzzi, and H. Dery,  Phys. Rev. B {\bf 88}, 045204 (2013). 
\bibitem{optical3} J. Lohrenz, T. Paschen, and M. Betz, Phys. Rev. B {\bf
    89}, 121201 (R) (2014).

\bibitem{Loren-Ge1} E.~J. Loren, B.~A. Ruzicka, L.~K. Werake, H. Zhao,
  H.~M. van Driel, and A.~L. Smirl, Appl. Phys. Lett. {\bf 95}, 092107 (2009).
\bibitem{Rioux-Ge1} J. Rioux and J. E. Sipe, Phys. Rev. B {\bf 81},
  155215 (2010).
\bibitem{Loren-Ge2} E.~J. Loren, J. Rioux, C. Lange, J.~E. Sipe, H.~M. van Driel,
  and A.~L. Smirl, Phys. Rev. B {\bf 84}, 214307 (2011).

\bibitem{Li-Si1} P. Li and H. Dery, Phys. Rev. Lett. {\bf 107}, 107203
  (2011).
\bibitem{Tang-Ge1} J. M. Tang, B. T. Collins, and M. E. Flatt\'{e},
  Phys. Rev. B {\bf 85}, 045202 (2012).

\bibitem{Jain-Ge1} A. Jain, C. Vergnaud, J. Peiro, J.~C. Le Breton, E. Prestat,
  L. Louahadj, C. Portemont, C. Ducruet, V. Baltz, A. Marty,
   {A. Barski, P. B. Guillemaud, L. Vila,
    J. P. Attan\'e, E. Augendre, H. Jaffr$\grave{\rm e}$s,
    J.~M. George, and M. Jamet,} 
  Appl. Phys. Lett. {\bf 101}, 022402 (2012).
\bibitem{Li-Ge2} P. Li, Y. Song, and H. Dery, Phys. Rev. B {\bf 86},
  085202 (2012).
 {\bibitem{Li-Ge3} P. Li, D. Trivedi, and H. Dery, Phys. Rev. B {\bf 87},
  115203 (2013).}


\bibitem{DP} M. I. D'yakonov and V. I. Perel', Zh. Eksp. Teor. Fiz. {\bf 60}, 1954
  (1971) [Sov. Phys. JETP {\bf 33}, 1053 (1971)].

\bibitem{purification1} B. E. Kane, Nature (London) {\bf 393}, 133 (1998).
\bibitem{purification2} P. S. Fodor and J. Levy, J. Phys.: Condens. Matter {\bf
    18}, S745 (2006).

\bibitem{Song_impurity} Y. Song, O. Chalaev, and H. Dery, Phys. Rev. Lett. {\bf 113}, 167201 (2014). 

\bibitem{interface} M. Tran, H. Jaffr\'es, C. Deranlot, J. M. George, A. Fert,
  A. Miard, and A. Lema\^itre, Phys. Rev. Lett. {\bf 102}, 036601 (2009).  
\bibitem{surface} S. P. Dash, S. Sharma, J. C. Le Breton, J. Peiro,
  H. Jaffr\'es, J. M. George, A. Lema\^itre, and R. Jansen, Phys. Rev. B {\bf 84},
  054410 (2011).

\bibitem{interface6} C. H. Li, O. M. J. van't Erve, and B. T. Jonker, Nature
Commun. {\bf 2}, 245 (2011).
\bibitem{interface7} S. P. Dash, S. Sharma, R. S. Patel, M. P. de Jong, and
R. Jansen, Nature (London) {\bf 462}, 491 (2009).

\bibitem{tunneling_im1} Y. Song and H. Dery, Phys. Rev. Lett. {\bf 113}, 047205 (2014).
\bibitem{tunneling_im2} O. Txoperena, Y. Song, L. Qing, M.
  Gobbi, L. E. Hueso, H. Dery, and F. Casanova, Phys. Rev. Lett. {\bf 113},
  146601 (2014).
\bibitem{tunneling_im4} H. N. Tinkey, P. Li, and I. Appelbaum, Appl. Phys. Lett. {\bf 104}, 232410 (2014).
\bibitem{tunneling_im3} A. G. Swartz, S. Harashima, Y.
  Xie, D. Lu, B. Kim, C. Bell, Y. Hikita, and H. Y. Hwang, Appl. Phys. Lett. {\bf 105}, 032406 (2014).
\bibitem{tunneling_im5} Z. Yue and M. E. Raikh, arXiv:1502.00350.

\bibitem{Yafet} Y. Yafet, Phys. Rev. {\bf 85}, 478 (1952).
\bibitem{Elliott} R. J. Elliott, Phys. Rev. {\bf 96}, 266 (1954).

\bibitem{negative} J. C. McGroddy, M. I. Nathan, and J. E. Smith, Jr IBM
  J. Res. DeV. {\bf  13}, 543 (1969).
\bibitem{hyperpure} C. Jacoboni, F. Nava, C. Canali, and G. Ottaviani, Phys.
Rev. B {\bf 24}, 1014 (1981).

\bibitem{wu-review} M. W. Wu, J. H. Jiang, and M. Q. Weng, Phys. Rep. {\bf 493},
   61 (2010).
\bibitem{jianhua15} M. W. Wu and H. Metiu, Phys. Rev. B {\bf 61}, 2945 (2000).
\bibitem{2001} M. W. Wu and C. Z. Ning, Eur. Phys. J. B. {\bf 18}, 373 (2000);
  M. W. Wu, J. Phys. Soc. Jpn. {\bf 70}, 2195 (2001).
\bibitem{jianhua23} M. Q. Weng and M. W. Wu, Phys. Rev. B {\bf 68}, 075312
  (2003).




\bibitem{Liu} Z. Liu, M. O. Nestoklon, J. L. Cheng, E. L. Ivchenko, and
  M. W. Wu, Fizika Tverdogo Tela {\bf 55}, 1510 (2013) [Phys. Solid State {\bf
    55}, 1619 (2013)].

\bibitem{Gamma} S. Ridene, K. Boujdaria, H. Bouchriha, and G. Fishman,
  Phys. Rev. B {\bf 64}, 085329 (2001).


\bibitem{jianhua52} H. Haug and A. P. Jauho, {\it Quantum Kinetics in Transport and
Optics of Semiconductors} (Springer, Berlin, 1996).



\bibitem{Yang_hot} E. M. Lifshitz and L. P. Pitaevskii, {\it Physical Kinetics} 
(Pergamon, London, 1981); G. E. Uhlenbeck, G. W. Ford, and E. W. Montroll,
{\it Lectures in Statistical Mechanics}
(American Mathematical
Society, Providence, 1963), Chap. IV; V. F. Gantmakher and
Y. B. Levinson,
{\it Carrier Scattering in Metals and Semiconductors}
(North-Holland, Amsterdam, 1987), Chap. 6.

\bibitem{Lei_hot} H. Fr\"ohlich and B. V. Paranjape, Proc. Phys. Soc. London,
Sect. B {\bf 69}, 21 (1956); K. Hess, in {\it Physics of Nonlinear Transport in Semiconductors},
edited by D. K. Ferry, J. R. Barker, and C. Jacoloni (Plenum,
New York, 1980), p.~1; D. K. Ferry, in {\it Physics of Nonlinear Transport in Semicon
ductors}, edited by D. K. Ferry, J. R. Barker, and C. Jacoloni
(Plenum, New York, 1980), p.~117; K. Seeger, {\it Semiconductor Physics} (Springer-Verlag, Berlin,
1982).

\bibitem{hot4}X. L. Lei, {\em Balance Equation Approach to Electron Transport in
  Semiconductors} (World Scientific, Singapore, 2008).



\bibitem{Jianhua} J. H. Jiang and M. W. Wu, Phys. Rev. B {\bf 79}, 125206 (2009);
  {\bf 83}, 239906(E) (2011).
\bibitem{Peng} P. Zhang, J. Zhou, and M. W. Wu,  Phys. Rev. B {\bf 77}, 235323 (2008). 
\bibitem{Hua} H. Tong and M. W. Wu, Phys. Rev. B {\bf 85}, 075203 (2012). 
\bibitem{Landolt}{\em Numerical Data and Functional Relationships in Science and
    Technology}, Landolt-B\"ornstein, New Series, Group III, Vol. 17, Pt. A,
  edited by O. Madelung, M. Schultz, and H. Weiss (Springer-Verlag, Berlin,
  1982). 


\bibitem{mobility} In the optical experiment by Lohrenz {\em et
    al.},\cite{optical3} the photogenerated peak
 carrier density is stated to be $1.5\times10^{13}$~cm$^{-3}$ by the optical
 pulse in the nominally undoped sample. In the electrical 
  experiment by Li {\em et
    al.},\cite{Li-Ge4} the samples are nominally undoped with the 
room-temperature resistivity $>40$ $\Omega\cdot$cm. As the mobility limited 
by the electron-phonon scattering at 300~K is $2.8\times 10^4$~cm$^2$/(V$\cdot$s),
  and hence  the corresponding electron density in the sample is 
fitted to be $<5\times 10^{12}$~cm$^{-3}$. However, this
density is not necessarily the one participating in transport in the 
 spin injection experiment by attaching magnetic contacts and
the injected electron density is not reported in
Ref.~\onlinecite{Li-Ge4}. Therefore, in the numerical calculation, we
  choose 
  the electron density to be $n_e=10^{13}$~cm$^{-3}$, which is a little higher than
  $5\times 10^{12}$~cm$^{-3}$.


\bibitem{Ka} K. Shen, Chin. Phys. Lett. {\bf 26}, 067201 (2009).
\bibitem{Ka_explain} M. Krau\ss{}, H. C. Schneider, R. Bratschitsch, Z. Chen,
  and S. T. Cundiff, Phys. Rev. B {\bf 81}, 035213 (2010).

\bibitem{Faraday_Kerr} N. Tombros, S. Tanabe, A. Veligura, C. J\'ozsa, M. Popinciuc, H. T. Jonkman, and B. J.
van Wees, Phys. Rev. Lett. {\bf 101}, 046601 (2008).
\bibitem{Awschalom} {\it Semiconductor Spintronics and Quantum Computation},
  edited by D. D. Awschalom, D. Loss, and N. Samarth (Springer, Berlin, 2002).
\bibitem{Aws_rotation2} J. A. Gupta, R. Knobel, N. Samarth, and D. D. Awschalom, Science {\bf 292}, 2458 (2001). 
\bibitem{Aws_rotation3} S. A. Wolf, D. D. Awschalom, R. A. Buhrman,
  J. M. Daughton, S. von Moln\'ar, M. L. Roukes, A. Y. Chtchelkanova, and 
  D. M. Treger, Science {\bf 294}, 1488 (2001).
\bibitem{Korn} T. Korn, Phys. Rep. {\bf 494}, 415 (2010).


\bibitem{hot_spot1} J. L. Cheng, M. W. Wu, and J. Fabian, Phys. Rev. Lett. {\bf 104}, 016601 (2010).
\bibitem{hot_spot2} P. Li and H. Dery, Phys. Rev. Lett. {\bf 107}, 107203 (2011).










\end{thebibliography}
\end{document}